# Clicks, comments, consequences: Are content creators' socio-structural and platform characteristics shaping the exposure to negative sentiment, offensive language, and hate speech on YouTube?

Sarah Weißmann[a]*, Aaron Philipp[a], Roland Verwiebe[a], Chiara Osorio Krauter[a], Nina-Sophie Fritsch[ab] and Claudia Buder[c]

[a] Faculty of Economics and Social Sciences, University of Potsdam, Potsdam, Germany
[b] Department of Sociology; University of Economics and Business, Vienna, Austria
[c] École Doctorale de Science politique, Panthéon-Sorbonne, Paris, France

*Corresponding author: sarah.weissmann@uni-potsdam.de

**Abstract**

Receiving negative sentiment, offensive comments, or even hate speech is a constant part of the working experience of content creators (CCs) on YouTube – a growing occupational group in the platform economy. This study investigates how socio-structural characteristics such as the age, gender, and race of CCs but also platform features including the number of subscribers, community strength, and the channel topic shape differences in the occurrence of these phenomena on that platform. Drawing on a random sample of n=3,695 YouTube channels from German-speaking countries, we conduct a comprehensive analysis combining digital trace data, enhanced with hand-coded variables to include socio-structural characteristics in social media data. Publicly visible negative sentiment, offensive language, and hate speech are detected with machine- and deep-learning methods using N=40,000,000 comments. Contrary to existing studies our findings indicate that female content creators are confronted with less negative communication. Notably, our analysis reveals that while BIPoC, who work as CCs, receive significantly more negative sentiment, they aren't exposed to more offensive comments or hate speech. Additionally, platform characteristics also play a crucial role, as channels publishing content on conspiracy theories or politics are more frequently subject to negative communication.

# Introduction

Content creators (CCs) are a crucial occupational group of cultural producers on social media platforms (Arriagada & Ibáñez, 2020; Craig & Cunningham, 2019) who 'pursue creative activities that hold the promise of social and economic capital' (Duffy, 2016, p. 443). A unique part of their professional life is the constant engagement with social media, the need to build up para-social relationships, and their ability to influence public opinion. In this interplay of fandom and strong support, constant feedback and often public scrutiny, CCs are exposed to negative comments, including offensive language, hate speech or even threats to their lives (Harris et al., 2023), with detrimental effects on their well-being (Vitak et al., 2017), life satisfaction (Stahel & Baier, 2023) and health (Heung et al., 2024). Negative communication, even if constructive, can threaten the CC's credibility and therefore impact the ability to promote products (Weber et al., 2024) and contribute to social media fatigue (Kwon et al., 2020). Because insults and hate are a daily occurrence, the majority of CCs employ coping strategies for this type of stress, even leading to the deletion of their accounts in some cases (Thomas et al., 2022). Existing studies indicate that CCs are not equally affected by negative communication depending on the gender and race or the channel size and the channel topic (Döring & Mohseni, 2020; Harris et al., 2023; Thomas et al., 2022; Vogels, 2021).

In this paper, we build upon key findings of existing research across various social media platforms and aim to extend it through three key aspects while focusing on content creators on YouTube: 1. *Comprehensive analysis of negative communication*: We examine different forms of negative communication patterns – specifically negative sentiment, offensive language, and hate speech – that content creators are exposed to. We aim for a clear analytical distinction, with negative sentiment reflecting the overall tone of a comment, offensive language indicating hostility and hate speech being directed specifically towards a particular group of people (for details, see section 3). By comparing and linking these forms, we strive to achieve a more thorough understanding of negative communication patterns within YouTube. 2. *Integration of contextual characteristics through multivariate analysis*: Utilizing a large random sample of YouTube channels, we conduct detailed investigations employing systematically socio-structural characteristics of content creators (e.g., age) with platform characteristics (e.g., audience size) for the first time, allowing us to elucidate their combined influence on negative communication dynamics. 3. *Broadening the scope beyond specific debates and topics*: Contrasting with previous studies that often focus on individual channels, selected niches, or rely on self-reports and qualitative data (Breazu & Machin, 2023; Wotanis & McMillan, 2014), our contribution facilitates a broader exploration of negative communication. This enables us to gain a more expansive and nuanced understanding of the various aspects and contexts in which negative interactions occur on YouTube allowing us to contribute to a theoretical discussion of new aspects of CCs' occupational praxis on algorithm-based markets (Barth et al., 2023). In this light, our research is guided by the following research question: How do socio-structural characteristics of content creators (e.g. age, gender, race, religion) in combination with platform features (e.g. channel topic, audience size, community strength) increase the risk of exposure to negative sentiment, offensive language, and hate speech on YouTube? To answer this question, we employ a unique dataset that combines digital trace data from 3,695 CCs on YouTube in German-speaking countries with socio-structural variables gathered through a hand-coded classification survey. This diverse sample of creators totals around 40 million publicly visible comments, which we analyze using machine learning and deep learning algorithms.

## State of the art

YouTube is one of the central markets in the platform economy, bringing together producers and consumers of cultural goods. In this industry, CCs operate as digital self-employed who publish their own content that potentially will be played out to a broader audience if it serves YouTube's business interests (Hoose & Rosenbohm, 2024). The reliance on the platform and its algorithmic structure fundamentally shapes and defines this emerging occupational field. Although CCs benefit from flexible working hours and the absence of a fixed workplace, they face continuous pressure to produce content, compounded by ever-evolving platform algorithms. This dynamic results in uncertain, and often precarious, working conditions characterized by diffused responsibility and limited proximity which significantly influence users' communication, behavior, and actions (Lowry et al., 2016). This new type of digital work, one could argue, resembles to some degree what Pongratz and Voß (2003, p. 243) in their seminal essay describe as 'entreployee'. This work requires 'self-determined organization, control and monitoring' of one's professional activities, 'intensified active and practical "production" and "commercialization" of one's own capacities' and 'the tendency to accept willingly the importance of the company' – in the present case YouTube as a platform – as an everyday integral part of one's own life (Pongratz & Voß, 2003, p. 44). In light of these specific circumstances, CCs are quite often striving to build a community, validate the meaningfulness of their work, and foster a relationship with their audience (Arriagada & Ibáñez, 2020; Bonifacio et al., 2023; Byun et al., 2023). While YouTube provides the opportunity for online engagement, enabling CCs and viewers to exchange opinions beneath the videos via comments, this also creates an inherent risk of being affected by negative communication (Obadimu et al., 2021).

Existing research indicates an inequality in the extent to which CCs are confronted with negative sentiment, offensive language, and hate speech (Blackwell et al., 2017; Feuston et al., 2020; Scheuerman et al., 2018). One main topic of relevance in this context is gender (Górska et al., 2023; Miyake, 2023; Shor et al., 2022). Eckert (2018) for example shows that female bloggers, who address politics, regularly experience various forms of online abuse. Correspondingly, Wotanis and McMillan (2014, p. 923) argue in their case study that female CCs on YouTube are often objectified in the comments, characterized by sexually explicit and offensive comments and even "supportive feedback consisting of compliments regarding […] physical appearance." Döring and Mohseni (2020) find for a comparison of eight channels that women on YouTube are confronted with more sexist comments in the form of degrading and benevolent stereotypes. While other studies report the occurrence of negative sentiment or offensive language and hate speech in female CCs comment sections for specific topics such as comedy (Döring & Mohseni, 2019a, 2019b), STEM (Amarasekara & Grant, 2019), and education (Veletsianos et al., 2018), comprehensive analyses beyond specific cases and selected channels is still rare.

In addition to gender, Park et al. (2021) identify age as another significant factor contributing to the occurrence of negative communication patterns. Specifically, age seems to influence the perception of hate speech with younger individuals detecting it more easily and older individuals tend to react to it more emotionally (Schmid et al., 2022). While the age of CCs as a target of hate speech is not of particular focus in much of the research, many studies tackle the influence, causes, occurrence, and prevention of hate speech specifically for adolescents using mostly subjective assessments (Kansok-Dusche et al., 2023; Obermaier & Schmuck, 2022). As adolescents use social media more actively (Bobzien et al., 2025; Harriman et al., 2020; Pew Research, 2023) and tend to engage in riskier online behavior than older adults (Koutamanis et al., 2015; Stahel &

Baier, 2023), the age of creators could be a significant risk factor for an increased exposure to negative communication patterns. However, as we are aiming at a deeper understanding of this phenomena, it remains unclear whether this trend, which is mainly observed through self-reported surveys, also displays itself in form of negative sentiment, offensive language, and hate speech in YouTube comments.

With ongoing discussions regarding the discriminatory nature of digital platforms and their reinforcement of racist dynamics (Matamoros-Fernández, 2017; McMillan Cottom, 2020), social media can create an environment, where race and religious affiliation emerge as significant risk factors for negative sentiment, offensive language, and hate speech (Castaño-Pulgarín et al., 2021; Haimson et al., 2021). For example, a study by Harris et al. reveals, based on 12 semi-structured interviews with African-American TikTok content creators, that CCs encounter "in particular anti-Black hate speech" (Harris et al., 2023, p. 16). However, the pervasive nature of discrimination extends beyond race; recent studies have found that religious discrimination especially affects Muslims and Jewish individuals both online and offline (Awan & Zempi, 2016; Ozalp et al., 2020; Weichselbaumer, 2020; Younes, 2020). This trend is also prevalent in the German context, where research indicates a general increase in anti-immigration and anti-refugee attitudes on social media (Aldamen, 2023; Paasch-Colberg et al., 2022). Beyond these general developments that primarily affect different user groups and the overall atmosphere on social media, studies for YouTube using quantitative data on whether and how CCs are confronted with negative sentiment, offensive language, and hate speech based on their race or religion are still relatively rare.

It's apparent that negative communication patterns we know from offline contexts, driven by individuals' socio-structural characteristics, are being replicated within the digital realm (Laor, 2022; Petters et al., 2024; Pew Research, 2023). However, interactions on social media introduce additional platform characteristics that can either hinder or facilitate the occurrence of various forms of negative communication. One driving factor for the distribution of content on social media platforms are algorithmic curation processes. This can push users towards extreme and misinforming videos (Bryant, 2020; Hussein et al., 2020; Yesilada & Lewandowsky, 2022) and enforces the emergence of filter bubbles and echo chambers (Cinelli, De Francisci Morales, et al., 2021; Diaz Ruiz & Nilsson, 2023) where users are "only presented with information that matches with [their] previous consumption behavior" (Spohr, 2017). At the same time, policy changes on YouTube (e.g., deplatforming, stricter monetization goals) negatively affected non-ad-friendly content especially harshly (Haimson et al., 2021; Kumar, 2019; Rauchfleisch & Kaiser, 2024). Another factor, according to ElSherief et al. (2018) is, that popular CCs with more followers are more often targets of negative communication patterns. In addition, it can be argued that existing comments shape further commenting behavior (Waddell & Bailey, 2017) leading sometimes to the emergence of even more toxicity (Cinelli, Pelicon, et al., 2021; Mathew et al., 2020) even as YouTube has continued using moderation tools against toxic content at multiple levels of governance.[1]

A key aspect of platform dynamics is the role of topics or specific niches. Research focusing on areas such as scientific knowledge or gaming often explores audience culture and composition (Salter, 2018) as well as underlying beliefs and biases present in these genres (Amarasekara & Grant, 2019) in relation to the occurrence of negative sentiment, offensive language, and hate speech. Vossen (2018) for example describes the existence of cultural inaccessibility for certain groups on gaming platforms, while Salter (2018) discusses 'geek masculinity' and its connection to online abuse (Díaz-Fernández & García-Mingo, 2024; Vergel et al., 2024). Thelwall et al. (2012) suggest that there is a difference in the communication cultures of certain topics, with music being a passive genre that is mostly only consumed, while politics had a much higher comment density. Another strand of literature focuses on the increasing hostility, violence, and populism of political communication on social media (Finlayson, 2022). Analyzing hate speech directed at American politicians on X shows that negative sentiment, offensive language, and hate speech are on the rise in the political sphere (Solovev & Pröllochs, 2022). In addition, the accessibility of YouTube for sharing user-generated content also appeals to conspiracy theorists. Numerous studies indicate that these phenomena are no longer isolated cases but have emerged as a significant topic, cultivating a distinct audience and specific communication patterns in comment sections (Allington et al., 2021; Shooman, 2016). Finally, there are content creators specifically inciting violence or hatred themselves as Stewart et al. (2023) show for Telegram, which can lead to a concentration of hate speech in digital environments. While these findings show that topic-specific negative communication patterns exist, certain commenters, as well as CCs, possibly self–select into these hate bubbles (Xin, 2024), where more hateful or negative communication logics apply. However, the existing literature has yet to address the extent to which such potential self-selection interacts with their exposure to negative sentiment, offensive or hate comments, particularly in relation to differences across topics and niches CCs choose for their YouTube channel.

---

[1] Up to 800 million videos are uploaded to YouTube each year. About 35 million videos were deleted by the company in 2024 (33.5 million through automated flagging). 55% were deleted due to child safety reasons, 8% for cyberbullying, 5% for sexual, 16% for harmful, and 9% for violent content (Google, 2025b).

Lastly, a key factor in the occurrence of negative communication in comment sections is who decides to engage in the first place. Generally, comments can be understood as a vehicle for externalizing emotional reactions (Alhabash et al., 2015; Krämer et al., 2021).

9The audience on social media platforms can be divided into a large number of passive users who only consume content and a small proportion of active users who comment, like, and share content (Cinelli, Pelicon, et al., 2021). Khan (2017), based on an online survey among YouTube users, is able to demonstrate that the male gender positively predicts disliking and commenting on videos. Furthermore, women seem to express positive emotions more frequently than men, as Sun et al. (2020) discuss with their study on a Chinese Python community. Differences in audiences, shaped by both viewing preferences or algorithmic curation, can therefore influence negative communication patterns. Audiences who consistently engage with content and possibly develop parasocial relationships with CCs (Lotun et al., 2022) tend to provide more favorable feedback and may even shield CCs from criticism through content moderation as Villegas-Simón et al. (2023) analyze with their qualitative study of 18 Spanish CCs from different social media platforms. Strong communities characterized by high levels of engagement through frequent commenting by the same people can foster positive communication cultures. We aim to systematically investigate this assumption using 40 million comments, considering other influencing factors, with a specific focus on YouTube.

While the research on these characteristics of content creators and platform related factors offers valuable insights into the various processes shaping the occurrence of negative sentiment, offensive language, and hate speech, the understanding of the interdependence of these various factors remains vague, especially for the German-speaking countries – the case under study. For our study, we bridge and systematize the current state of research, resulting in our hypotheses illustrated in Figure 1. They form the basis for our subsequent empirical analyses of how socio-structural characteristics of content creators, in combination with platform features, increase the risk of exposure to negative sentiment, offensive language, and hate speech on YouTube.

Figure 1: Illustration of the relationship between socio-structural and platform characteristics and negativity, offensive language and hate speech

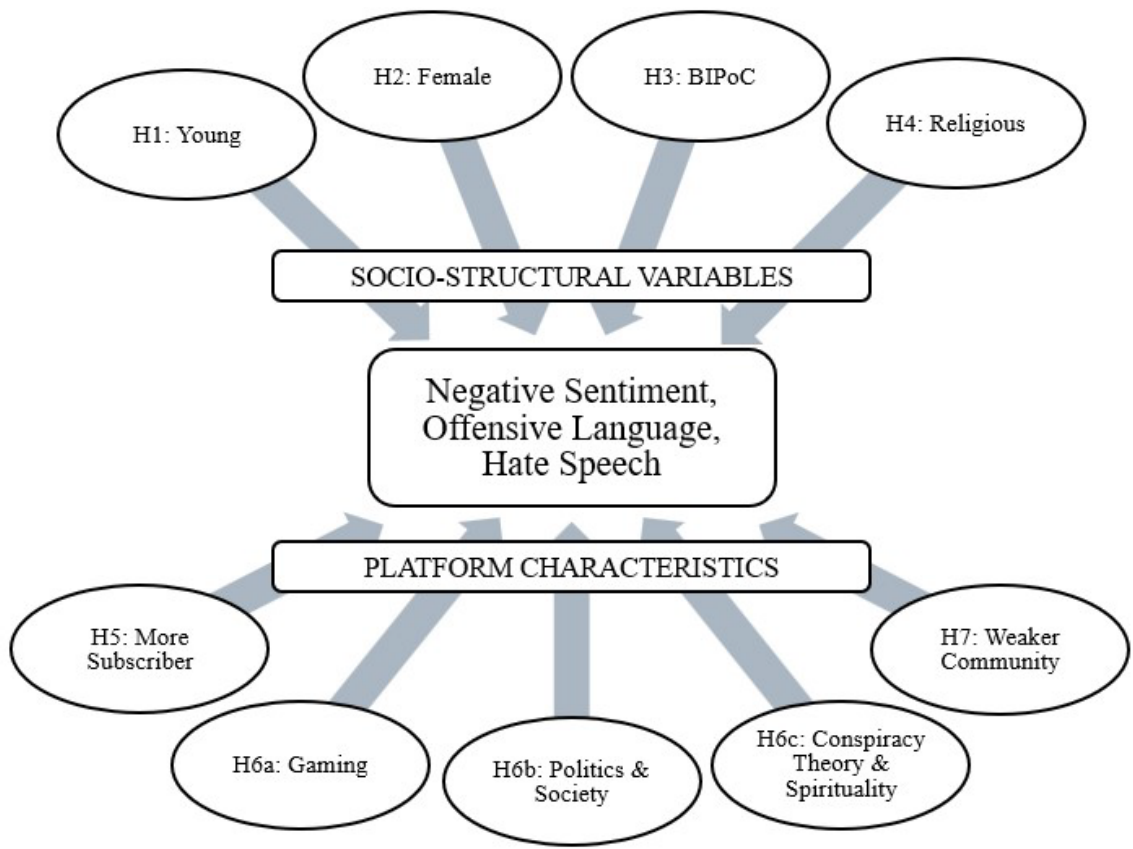

## Data and methods

**Data**

The data for the empirical analysis is drawn from the population of N=115,975 channels registered in Germany, Austria, Switzerland, or Liechtenstein, each of which has uploaded at least 10 videos to YouTube, which we obtained through the website

channelcrawler.com in December 2022.[2] Because the present contribution is focused on CCs, channels hosted by companies, government agencies, political parties, or NGOs are excluded. Furthermore, we restrict the sample to CCs with mainly German or English[3] comments. Our final sample consists of 3,695 channels, N ≈ 430,000 videos, and N ≈ 40,000,000 publicly visible comments. Our data is composed of (1.) platform characteristics such as the number of subscribers, accumulated views, and comments received and (2.) socio-structural characteristics of CCs including age, gender, race, religious affiliation.

(1.) Between July and December 2023, we compiled the channel information alongside platform metrics including the number of subscribers, views, likes, and publication data and collected all comments and replies under the videos of our 3,695 channels using the YouTube Data API v3.

(2.) To fill the lack of missing socio-structural variables in digital trace data, we developed a standardized classification survey (Liang et al., 2022; Seewann et al., 2022) and hand-annotated age, gender, race, religious affiliation, and the channel topic of the CCs employing six coders.[4] The classification survey included (1) basic information such as the profile picture, channel description, and various platform metrics, (2) the most recently uploaded video to offer additional information, and (3) the social media links obtained from the CCs' profiles allowing coders further online research to gather even more comprehensive information.

---

[2] Channelcrawler provided us with the channel ID that we used to scrape further information via the YouTube API.
[3] We included English comments because social media communication in Germany (and many other countries outside of the U.S. or UK) often incorporates English terms (e.g., 'sick,' 'epic,' 'nice').
[4] To measure the intercoder reliability we used Fleiss' Kappa, which resulted in 0.76 for gender and 0.58 for age.

*Dependent variables*

We use the XGBoost algorithm for predicting sentiment (Liu, 2020) and a multilanguage BERT model[5] for the detection of publicly visible offensive language and hate speech[6] (Jahan & Oussalah, 2023) for our N ≈ 40,000,000 comments.[7] The training data we employ for these models is based on 7,500 German and English YouTube comments that are manually annotated regarding their sentiment, the occurrence of offensive language and hate speech (Kenyon-Dean et al., 2018; Medhat et al., 2014). This dataset and further information on the annotation is available on https://github.com/Sarahanna/Hate-speech-and-sentiment-classification-and-dictionary.

Sentiment prediction was performed by applying the following values to each comment: 1 for positive, 0 for neutral, and -1 for negative. Offensive language and hate speech were detected using a categorical approach, where 2 indicated the occurrence of hate speech, 1 offensive language and 0 the absence of both.[8] After predicting the sentiment and the occurrence of offensive language or hate speech for each comment we aggregated the data into three dependent variables using all comments under the channel's videos. Therefore, on channel level we calculated (1) the mean sentiment (2) the proportion of comments containing offensive language, and (3) the proportion of comments containing hate speech.

---

[5] We tested additional methods for all three independent variables: decision tree, multinomial logistic regression, random forest, support vector machine. The XGBoost algorithm (macro F1: 0.60) for sentiment and the BERT model (macro F1: 0.69) for offensive language and hate speech achieved the highest performance on our dataset.

[6] Hate speech detection is often applied under several broad terms like toxicity, harassment, hate, or offensiveness (Jahan & Oussalah, 2023). Our study investigates two specific concepts that fit within the broader field of hate speech detection. We extended the Code of Conduct between the EU and IT companies (European Commission, 2016) by further including gender, sexual orientation, political identity, and false allegations and ultimately defined hate speech as 'Any behavior that incites violence or hatred against individuals or a group of individuals or a member defined by reference to race, color, religion, descent or national or ethnic origin, gender, sexual orientation, political opinion or makes false allegations.' On the other hand, we defined offensive language as 'Comments which are insulting, toxic or hostile but are not exclusively directed towards protected groups.'

[7] The data cleaning process involved converting all text to lowercase, removing website links and hashtags, and recoding emojis into text. Additionally, for XGBoost, punctuation and stop words were removed.

[8] Offensive comments include hate speech since they are also a form of offensive comments. Robustness checks, which excluded hate from offensive comments revealed no systematic differences.

*Independent variables*

We include key predictors for negative communication patterns identified by the existing research. Age is measured as a categorical variable, starting with 'under 20 years' and increasing in ten-year intervals, with the final category being '40 years and above', as it could not be captured on a metric scale. Gender and religious affiliation are coded as dichotomous variables and race is recoded from a five-scale ordinal variable to 0 = white and 1 = BIPoC. All socio-structural variables have a 'mixed' category for group channels consisting of different demographic groups, e.g., male and female hosts. Community strength measures the number of recurring commenters on a channel, scaled from 0 to 1, with higher values indicating a stronger community. This metric variable as well as the subscriber count were standardized for the analysis. The channel topic consists of 14 categories. Controls include the channel age and whether a channel is monetizing its content through YouTube (YouTube, 2024). The quality of information varies across the different variables, as one can see in Table 1. The sample is predominantly male (70%) and white (50%), with 85% of participants lacking an identifiable religious affiliation. Regarding the platform characteristics, we find an imbalanced distribution in the number of subscribers (mean = 24,744; median = 524) and varying sizes of the topics (gaming = 1,280; politics & society = 24), with the topic indicating the channel's thematic focus.

Examining the distribution of comments across various topics, we observe an average of 6,000 comments per video in DIY (median = 860), while political channels reach up to an average of 35,000 comments (median = 1,000). Videos in Arts & Culture, with comment counts ranging from 2 to 2,611,105 highlights the significant variability and skewness in commenting behaviour across different topics.

Table 1: Sample composition

| Socio-structural variables | N | % | Platform characteristics | Mean, N | % |
|---|---|---|---|---|---|
| Channel hosted by | | | Community Strength ∈ [0, 1] | | |
|    singles | 3,490 | 94.5 |    Min | 0 | |
|    groups | 205 | 5.5 |    Mean (SD) | 0.53 (0.23) | |
| Age | | |    Median | 0.55 | |
|    ≤ 20 years | 546 | 14.9 |    Max | 0.99 | |
|    21-30 years | 809 | 21.9 | Channel age [in years] | | |
|    31-40 years | 578 | 15.6 |    Min | 1.75 | |
|    40+ years | 656 | 17.7 |    Mean (SD) | 9.59 (3.61) | |
|    Mixed | 33 | 0.9 |    Median | 9.17 | |
|    not identified | 1,073 | 29.0 |    Max | 18.42 | |
| Gender | | | Subscribers | | |
|    Female | 558 | 15.1 |    Min | 9 | |
|    Male | 2,571 | 69.6 |    Mean (SD) | 24,744 (314,539) | |
|    Mixed | 75 | 2.0 |    Median | 524 | |
|    not identified | 491 | 13.3 |    Max | 15,800,000 | |
| Race | | | Monetization | | |
|    BIPoC | 345 | 9.4 |    Yes | 702 | 19.0 |
|    White | 1,818 | 49.2 |    No | 2,993 | 81.0 |
|    Mixed | 13 | 0.3 | Channel topic | | |
|    not identified | 1,519 | 41.1 |    Arts & Culture | 478 | 12.9 |
| Religious affiliation | | |    Beauty & Lifestyle | 121 | 3.3 |
|    Yes | 26 | 0.7 |    Business & Finances | 39 | 1.1 |
|    No | 525 | 14.2 |    Conspiracy Theory & Spirituality | 94 | 2.5 |
|    Mixed | 0 | 0.0 |    DIY | 300 | 8.1 |
|    not identified | 3,144 | 85.1 |    Education & Knowledge | 95 | 2.6 |
| | | |    Entertainment | 807 | 21.8 |
| | | |    Food & Culinary | 71 | 1.9 |
| | | |    Gaming | 1,280 | 34.6 |
| | | |    Health | 77 | 2.1 |
| | | |    Politics & Society | 24 | 0.4 |
| | | |    Sport | 119 | 0.6 |
| | | |    Travel | 174 | 3.2 |
| | | |    Other | 16 | 4.7 |
| Observations | | | | 3,695 | |

**Limitations**

While platforms like X and Meta have retracted from implementing platform moderation and measures against hate speech and fake news (BBC, 2025), YouTube has continued using several tools against problematic content (Google, 2025b; Jhaver & Zhang, 2023; YouTube, 2019). These shape our outcomes in several ways: Overall exposure to negative sentiment, offensive language, and hate speech is likely lowered due to the platform's algorithms and human moderators, which filter both videos and comments. Furthermore, YouTube provides tools for CCs, such as deciding the level of automatic filtering (basic, strict), word filter tools, blocking and reporting of users, which is also practiced by the viewers watching content, and deletion of comments (Google, 2025a, 2025b). While this could result in a more positive publicly visible comment section, the specific extent to which content moderation occurs on the level of individual channels remains largely unknown (Dergacheva & Katzenbach, 2023) and is thus difficult to consider in statistical analyses[9] (see the next section for empirical tendencies on CCs moderation in this study).

**Results**

In a first descriptive analysis of the N=40 million cases, we looked at the occurrence of publicly visible negative communication patterns on the comment level, based on the machine learning models at hand. The sentiment analysis shows a positive prevalence of 16.41%, a neutral prevalence of 75.82%, and a negative prevalence of 7.76%, indicating that most comments are neutral. The occurrence of negative comments is less than half as prominent as positive comments. In comparison, harmful language is relatively rare, with 2.73% of comments predicted as offensive and 0.83% as hate speech, indicating that the vast majority of comments were neither offensive nor hateful.[10]

Looking first at the overall tone of the comment section, the results of the OLS regressions in Fig. 2 show that CCs on YouTube are unequally exposed to negative sentiment.

---

[9] As a result, most quantitative studies on these topics fail to account for both individual moderation by CCs and the platform's algorithmic governance. They either worked with self-reported data (Aldamen, 2023; Eckert, 2018), merely state moderation as a limitation (Döring & Mohseni, 2020; Veletsianos et al., 2018) or don't mention content moderation as a relevant factor at all (Allington et al., 2021; Cinelli, De Francisci Morales, et al., 2021).

[10] These numbers relate to visible comments after potential moderation. In an unpublished survey among N=480 CCs in Germany, we asked some questions on their weekly moderation routines. According to the participants of this survey, they delete 9 comments on average per week, highly educated CCs delete more comments, women delete less hate comments then men and CCs with a political channel, entertainment channel, or gaming channel delete more comments than CCs with topics such as DIY, cooking, sport/fitness.

16Starting with the hypotheses related to socio-structural variables, we find that female CCs receive significantly more positive sentiment compared to men (0.058, p<.001) *(H1)*.[11] While this indicates the relevance of gender for the occurrence of certain communication patterns, this leads to a rejection of our hypothesis. Furthermore, we find a clear pattern for age *(H2)*: Younger CCs are exposed to significantly more negative publicly visible comments compared to the reference category of 40+ years CCs. Adolescents under the age of 20 (-0.045, p<.001) and 21-30 years old CCs (-0.052, p<.001) are especially affected by a more negative environment regarding the sentiment in their comment section. Furthermore, there is evidence supporting *H3*, as significant associations were found between race and sentiment with BIPoC content creators being exposed to more negative sentiment than channel hosts who are white (-0.038, p<.001). There is no significant evidence for *H4* regarding religious affiliation.
The platform variables are relevant as well: the number of subscribers, the channel topic, and the community strength show significant effects. An increase in subscribers causes a decrease of positive comments (-0.005, p<.05), leading to more negative sentiment in the comment section *(H5)*. Compared to the reference category arts, the topics DIY (0.046, p<.001), sport (0.045, p<.001) and travel (0.077, p<.001) are associated with a higher occurrence of positive sentiment. In contrast the channel topics conspiracy (-0.064, p<.001) and politics (-0.086, p<.01) exhibit significant negative coefficients indicating evidence for *H6*. In addition, we observe a significant coefficient for community strength (0.029, p<.001) *(H7)*, indicating that the closeness of a CC's online community is positively associated with positive sentiment.

---

[11] Each hypothesis is tested using the full model controlled for all variables (see Tab. 3).

Figure 2: OLS Regression of Sentiment

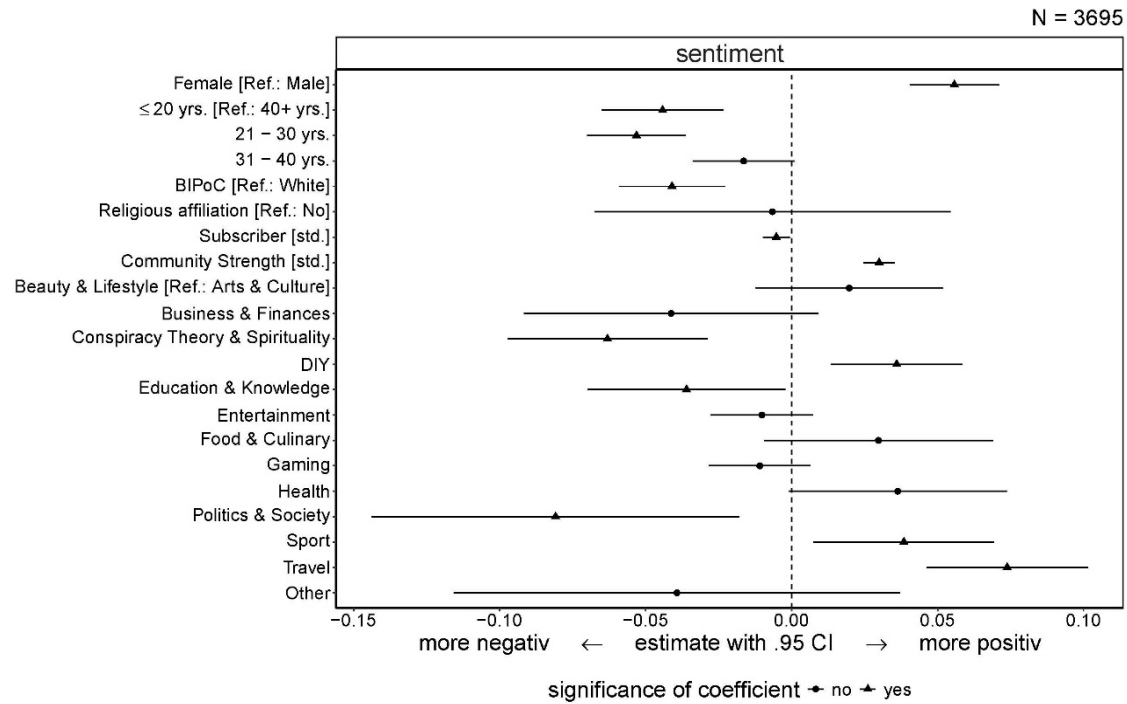

The plots in Fig. 3 show two separate OLS regressions for the influence of our independent variables on offensive language and hate speech. Similarly to sentiment and contradictory to our hypothesis, women are exposed to significantly fewer publicly visible offensive comments (-0.009, p<.001) and also less hate speech (-0.002, p<.01) than men *(H1)*. Our findings reveal an age effect (*H2)*, consistent with our assumptions, indicating that younger individuals are exposed to significantly more offensive language (0.005, p<.05 for CCs younger than 20 and 0.006, p<.01 for CCs between 21-30 years), while there is no significant effect for 31-40 years old CCs. We do not observe age affecting the occurrence of hate speech that CCs are exposed to. We find no indication of increased offensive language or hate speech related to race *(H3)* or religious affiliation *(H4)*. Focusing on platform characteristics, there is no evidence to support *(H5)*. The number of subscribers is not affecting the occurrence of offensive language or hate speech in the comment section. The topics knowledge (offensive: 0.012, p<.01; hate: 0.007,

p<.001) and entertainment (offensive 0.008, p<.001; hate: 0.002, p<.01) increase the probability of offensive language and hate speech on the channel. Meanwhile, gaming channels (0.004, p<.05) are associated with a higher occurrence of offensive comments only. Moreover, both offensive comments and hate speech are notably structured by the topics conspiracy as well as politics. Both topics have a significantly higher occurrence of offensive comments (conspiracy: 0.046, p<.001; politics: 0.065, p<.001) and hate speech (conspiracy: 0.021, p<.001; politics: 0.031, p<.001). This largely supports *H6*, demonstrating a strong association between controversial topics and hate speech, while gaming provides only weak evidence for this link. Lastly, with an increase in community strength, the occurrence of offensive comments (-0.005, p<.001) and hate speech (-0.0005, p<.05) decreases, supporting *H7*.

Figure 3: OLS Regressions of Offensive Language and Hate Speech

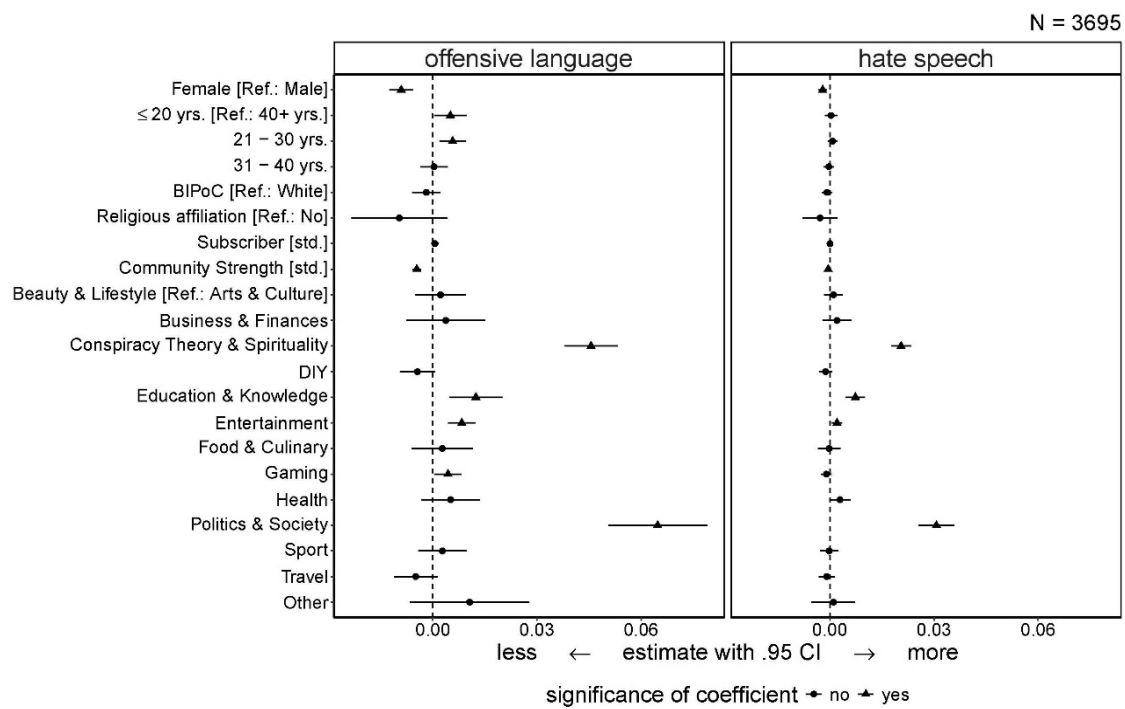

19## Discussion and conclusion

The exposure to comments marks a unique characteristic of CCs on YouTube, a growing occupational group within algorithm-based platforms. As digital 'entreployees' (Pongratz & Voß, 2003), their work requires not only increased individualized responsibility, self-determined organization and intensified commercialization of their own professional activities as an essential part of their life. With social media platforms serving as their working environment, CCs' heightened visibility and frequent interaction with their audience makes this occupation, unlike most other professions, especially vulnerable to negative communication (Dergacheva & Katzenbach, 2023). Developing protective strategies and even coping with the social and psychological consequences have become integral aspects of their daily work experiences (Heung et al., 2024; Thomas et al., 2022). Looking at content creators on YouTube in German-speaking countries, CCs received, on average, 9,345 comments per channel that are publicly visible. Among these, based on our estimations, they are exposed to hate speech about 100 times, offensive language 257 times, and 775 comments with negative sentiment since the foundation of their respective YouTube presence.[12] Although the overall sentiment across all channels is generally neutral, positive sentiment outweighs the negative. These numbers highlight that, despite an overall positive trend, negative communication remains a significant challenge for the work of CCs on YouTube.

Several central points can be summarized, directly linking back to the key arguments outlined in the introduction of this paper:

---

[12] Using negative sentiment as an example, these numbers translate to an average of 8 negative comments per month, with a standard deviation of 78. Two creators, one in health and the other in politics, even received a maximum of 3,282 and 1,478 negative comments in a single month, respectively.

(1) Our comparing analyses of sentiment, offensive language, and hate speech enable us to conduct a detailed, fine-grained study capturing nuances from emotional expressions to discriminatory content in YouTube comments. This approach broadens the understanding of negative communication structures within YouTube and uncovers some results that would otherwise have remained undetected. Through a systematic comparison of negative sentiment, offensive language, and hate speech, we are able to quantitatively assess and measure the dimensions of these three phenomena in terms of frequency and context of a CCs professional experience, as they appear on the platform. We also now know – referring directly to our research question at this point – that all three phenomena are stratified both by the social-structural composition of CCs and by the characteristics of the platform. Statistically, platform variables contribute more significantly to explaining the variation in our dependent variables, which can be interpreted as an indication of the high relevance of the platform's algorithmic structure (Bandy, 2021; Bishop, 2019). Another more specific example are the different results regarding the race of CCs. While CCs, who are BIPoC, are exposed to significantly more publicly visible comments with negative sentiment, they're not confronted with a higher occurrence of offensive language or hate speech in the case of the German-speaking countries under study. This shows that CCs are addressed differently in online spaces and that BIPoCs might face disadvantages even if content moderation systems are working.

(2) Combining digital trace data and annotated socio-structural variables allows researchers to provide a more comprehensive, larger-scale quantitative analysis of social media data. We were able to investigate whether individual characteristics contribute to the formation of at-risk groups among CCs and reveal that negativity and hate on YouTube do not appear at random. Instead, there are identifiable factors that influence the extent and severity of exposure to such comments, each contributing to a deeper understanding of negative online communication while accounting for one other.

For socio-structural characteristics, we found that specifically gender and age seem to structure sentiment, offensive language, and hate speech. Counterintuitively to the public perception, but also previous studies (KhosraviNik & Esposito, 2018), women are exposed to more positive sentiment and less offensive language or hate speech in their publicly visible comments (after controlling for the effects of age, race, channel topic etc.). One explanation could be systematic differences in audience composition, which may lead to women experiencing more exposure to positive online behavior. Since women are found to comment more positively than men (Sun et al., 2020) and are, on top of that, likely to exhibit homophily in online spaces (Pignolet et al., 2024), the comment section of female CCs could be more positively toned.[13] In addition, younger CCs were exposed to more negative sentiment and offensive language, maybe also due to a younger audience that encounters high social media use and shows riskier behavior (Koutamanis et al., 2015; Stahel & Baier, 2023), but not to more hate speech. It should be noted that age isn't included as a particular group in our definition of hate speech (age-related insults are thus categorized as offensive).

Regarding platform characteristics, the significant effects of subscribers and community strength on the various forms of negative communication we studied in this paper underline the relevance of the audience. We presumed the popularity of channels to be positively associated with the occurrence of the examined patterns (ElSherief et al., 2018). While this is the case for sentiment, the effect disappears for offensive comments and hate speech when including community strength. It indicates that strong community bonds can mitigate negative interactions and play a protective role against harmful comments (Lotun et al., 2022). This highlights the potential of effective community management and audience relationship-building as a powerful tool for CCs.

---

[13] It is a key topic for future research whether CCs, who are at risk of experiencing more negative communication on social media (e.g., BIPoCs (Harris et al., 2023), CCs with a large audience, women (KhosraviNik & Esposito, 2018), are also engaging into deleting and filtering more comments. In our unpublished survey, we could not find this result for women.

(3) The advantage of studying the platform across existing topics and forms of negative communication allowed us to assess what topics are especially affected. For example, gaming channels were presumed to be exposed to more negative sentiment, offensive language or hate speech than other channels due to the communities' specific communication culture which is at least partially reflected in our empirical findings (Salter, 2018). Moreover, topics that include potentially controversial discussions stand out clearly with their occurrence of hate speech (e.g. political, educational, or science content). Contentious topics are especially likely to attract heated conversations, and 'alternative facts' or political conversations attract people with controversial opinions that presumably view content moderation critically. The algorithmic structure of YouTube can further reinforce these patterns (Yesilada & Lewandowsky, 2022). It is assumed that opinion-based homophily is facilitated by certain social media platforms, leading to the formation of groups that inhibit specific hate-based communication (Evolvi, 2019). This reflects some of the previous work around hate bubbles or echo chambers which are formed based on similar beliefs but even go as far as forming shared identities (Nguyen, 2020; Xin, 2024).

**Limitations**

There are several limitations to this study. (1.) Regarding the data annotation of socio-structural variables, there is uneven access to information: Gender is a relatively reliable variable while education or religious affiliation are significantly more challenging to ascertain on YouTube. Furthermore, the annotation of sensitive personal information like race must be conducted and reflected upon with the utmost care. (2.) The NLP techniques used in this study face challenges with special linguistic features such as irony, sarcasm, or sexism which limits the predictive power for some comments, especially considering the highly dynamic nature of internet communication (Davidson et al., 2017; Ravi & Ravi, 2015). (3.) YouTube offers an extensive catalogue of tools to moderate comments, from automated hate speech detection to customizable word filters, and even allows the involvement of audiences by flagging content. However, there is only limited knowledge on the specific amount of content moderation on the level of individual YouTube channels that could be utilized for the statistical analysis. Therefore, statements about the absolute level of negative communication on YouTube should be made with caution, as both the public and we, as a scientific research group, can only analyze comments after moderation. Nonetheless, examining this issue remains highly relevant, as the risk of being exposed to visible negative communication varies significantly between different creators. (4.) Lastly, while it is known, that both increasing user engagement (Spinelli & Crovella, 2020) and remaining advertiser-friendly (Ma & Kou, 2021) is part of YouTube's business interest and therefore impact the function of the algorithm, we do not fully understand the platform's algorithmic behavior: Both the recommendation algorithm, which distributes the content presented to the viewers, and YouTube's hate speech detection algorithm, automatically filter community guideline violations are part of the algorithm 'black box' (Bishop, 2019) and are not accessible for our research or that of other scholars in the community.

# Appendix

## Appendix 1: OLS Regressions of Sentiment on Channel level

| | Dependent Var.: Sentiment on Channel Level | | |
|---|---|---|---|
| | Socio-structure | Platform characteristics | Full model |
| Gender [Ref.: Male] | | | |
|   Female | 0.062*** (0.007) | | 0.058*** (0.008) |
|   Mixed | 0.058** (0.020) | | 0.035 (0.020) |
|   Not identified | – 0.024* (0.009) | | – 0.016 (0.009) |
| Age [Ref.: 40+ years] | | | |
|   ≤ 20 years | – 0.045*** (0.010) | | – 0.045*** (0.011) |
|   21-30 years | – 0.056*** (0.008) | | – 0.052*** (0.009) |
|   31-40 years | – 0.021* (0.009) | | – 0.016 (0.009) |
|   Mixed | – 0.051 (0.030) | | – 0.038 (0.029) |
|   Not identified | – 0.053*** (0.011) | | – 0.047*** (0.011) |
| Race [Ref.: White] | | | |
|   BIPoC | – 0.054*** (0.009) | | – 0.038*** (0.009) |
|   Mixed | 0.028 (0.046) | | 0.0002 (0.044) |
|   Not identified | – 0.026** (0.008) | | – 0.021** (0.008) |
| Religious affiliation [Ref.: No] | | | |
|   Yes | – 0.026 (0.032) | | – 0.012 (0.031) |
|   Not identified | 0.002 (0.008) | | 0.003 (0.008) |
| Community Strength [std.] | | 0.034*** (0.003) | 0.029*** (0.003) |
| Subscriber [std.] | | – 0.005* (0.002) | – 0.005* (0.002) |
| Channel Topic [Ref.: Arts & Culture] | | | |
|   Beauty & Lifestyle | | 0.057*** (0.016) | 0.026 (0.016) |
|   Business & Finances | | – 0.027 (0.026) | – 0.033 (0.026) |
|   Conspiracy Theory & Spirituality | | – 0.045* (0.018) | – 0.064*** (0.017) |
|   DIY | | 0.056*** (0.012) | 0.046*** (0.011) |
|   Education & Knowledge | | – 0.018 (0.018) | – 0.029 (0.017) |
|   Entertainment | | – 0.015 (0.009) | – 0.009 (0.009) |
|   Food & Culinary | | 0.070*** (0.020) | 0.035 (0.020) |
|   Gaming | | – 0.034*** (0.009) | – 0.016 (0.009) |
|   Health | | 0.074*** (0.019) | 0.041* (0.019) |
|   Politics & Society | | – 0.085** (0.033) | – 0.086** (0.032) |
|   Sport | | 0.050** (0.016) | 0.045** (0.016) |
|   Travel | | 0.088*** (0.014) | 0.077*** (0.014) |
|   Other | | – 0.030 (0.040) | – 0.042 (0.039) |
| Constant | 0.223*** (0.009) | 0.183*** (0.007) | 0.219*** (0.011) |
| Observations | 3,695 | 3,695 | 3,695 |
| R2 | 0.071 | 0.099 | 0.140 |
| Adjusted R2 | 0.068 | 0.095 | 0.133 |
| Residual Std. Error | 0.158 (df = 3681) | 0.156 (df = 3677) | 0.153 (df = 3664) |
| F Statistic | 21.777*** (df = 3681) | 23.852*** (df = 3677) | 19.920*** (df = 3664) |

*Controlled for: monetization; channel age*     * $p < .05$; ** $p < .01$; *** $p < .001$

Appendix 2: OLS Regressions of Offensive Comments on Channel level

|  | Dependent Var.: Offensive Language on Channel Level | | |
|---|---|---|---|
|  | Socio-structure | Platform characteristics | Full model |
| Gender [Ref.: Male] | | | |
|   Female | – 0.008*** (0.002) | | – 0.009*** (0.002) |
|   Mixed | – 0.007 (0.005) | | – 0.003 (0.005) |
|   Not identified | 0.006** (0.002) | | 0.004 (0.002) |
| Age [Ref.: 40+ years] | | | |
|   ≤ 20 years | 0.002 (0.002) | | 0.005* (0.002) |
|   21-30 years | 0.003 (0.002) | | 0.006** (0.002) |
|   31-40 years | – 0.001 (0.002) | | 0.0003 (0.002) |
|   Mixed | 0.001 (0.002) | | 0.001 (0.007) |
|   Not identified | 0.008** (0.002) | | 0.008** (0.002) |
| Race [Ref.: White] | | | |
|   BIPoC | 0.0003 (0.002) | | – 0.002 (0.002) |
|   Mixed | – 0.007 (0.010) | | – 0.005 (0.010) |
|   Not identified | 0.0007 (0.002) | | 0.0003 (0.002) |
| Religious affiliation [Ref.: No] | | | |
|   Yes | – 0.001 (0.007) | | – 0.010 (0.007) |
|   Not identified | 0.001 (0.002) | | 0.001 (0.002) |
| Community Strength [std.] | | – 0.005*** (0.001) | – 0.005*** (0.001) |
| Subscriber [std.] | | 0.001 (0.001) | 0.001 (0.001) |
| Channel Topic [Ref.: Arts & Culture] | | | |
|   Beauty & Lifestyle | | – 0.005 (0.004) | 0.002 (0.004) |
|   Business & Finances | | 0.003 (0.006) | 0.003 (0.006) |
|   Conspiracy Theory & Spirituality | | 0.044*** (0.004) | 0.046*** (0.004) |
|   DIY | | – 0.005 (0.003) | – 0.005 (0.003) |
|   Education & Knowledge | | 0.011** (0.004) | 0.012** (0.004) |
|   Entertainment | | 0.010 (0.002) | 0.008*** (0.002) |
|   Food & Culinary | | – 0.003 (0.004) | 0.003 (0.005) |
|   Gaming | | 0.008*** (0.002) | 0.004* (0.002) |
|   Health | | 0.0003 (0.004) | 0.005 (0.004) |
|   Politics & Society | | 0.064*** (0.007) | 0.065*** (0.007) |
|   Sport | | 0.002 (0.004) | 0.003 (0.004) |
|   Travel | | – 0.005 (0.003) | – 0.005 (0.003) |
|   Other | | 0.010 (0.009) | 0.011 (0.009) |
| Constant | 0.018*** (0.002) | 0.016*** (0.002) | 0.012*** (0.002) |
| Observations | 3,695 | 3,695 | 3,695 |
| R2 | 0.030 | 0.095 | 0.116 |
| Adjusted R2 | 0.027 | 0.091 | 0.109 |
| Residual Std. Error | 0.036 (df = 3681) | 0.035 (df = 3677) | 0.035 (df = 3664) |
| F Statistic | 8.796***(df = 3681) | 22.652*** (df = 3677) | 16.001*** (df = 3664) |
| *Controlled for: monetization; channel age* | | | * p < .05; ** p < .01; *** p < .001 |

Appendix 3: OLS Regressions of Hate Speech on Channel level

| | Dependent Var.: Hate Speech on Channel Level | | |
|---|---|---|---|
| | Socio-structure | Platform characteristics | Full model |
| Gender [Ref.: Male] | | | |
|     Female | – 0.001 (0.001) | | – 0.002** (0.001) |
|     Mixed | – 0.0004 (0.002) | | 0.00002 (0.002) |
|     Not identified | 0.004*** (0.001) | | 0.003*** (0.001) |
| Age [Ref.: 40+ years] | | | |
|     ≤ 20 years | – 0.002* (0.001) | | 0.0003 (0.001) |
|     21-30 years | – 0.001 (0.001) | | 0.001 (0.001) |
|     31-40 years | – 0.001 (0.001) | | – 0.0003 (0.001) |
|     Mixed | – 0.001 (0.003) | | – 0.001 (0.002) |
|     Not identified | – 0.001 (0.001) | | 0.0003 (0.001) |
| Race [Ref.: White] | | | |
|     BIPoC | – 0.0003 (0.001) | | – 0.001 (0.001) |
|     Mixed | – 0.001 (0.004) | | – 0.002 (0.004) |
|     Not identified | – 0.0002 (0.001) | | 0.0001 (0.001) |
| Religious affiliation [Ref.: No] | | | |
|     Yes | 0.001 (0.003) | | – 0.003 (0.003) |
|     Not identified | – 0.0001 (0.001) | | 0.0001 (0.001) |
| Community Strength [std.] | | – 0.001*** (0.0002) | – 0.0005* (0.0002) |
| Subscriber [std.] | | 0.00005 (0.0002) | 0.00004 (0.0002) |
| Channel Topic [Ref.: Arts & Culture] | | | |
|     Beauty & Lifestyle | | – 0.001 (0.001) | 0.001 (0.001) |
|     Business & Finances | | 0.002 (0.002) | 0.002 (0.002) |
|     Conspiracy Theory & Spirituality | | 0.020*** (0.001) | 0.021*** (0.001) |
|     DIY | | – 0.001 (0.001) | – 0.001 (0.001) |
|     Education & Knowledge | | 0.007*** (0.001) | 0.007*** (0.001) |
|     Entertainment | | 0.002** (0.001) | 0.002** (0.001) |
|     Food & Culinary | | – 0.002 (0.002) | – 0.0002 (0.002) |
|     Gaming | | – 0.001 (0.001) | – 0.001 (0.001) |
|     Health | | 0.002 (0.002) | 0.003 (0.002) |
|     Politics & Society | | 0.031*** (0.003) | 0.031*** (0.003) |
|     Sport | | – 0.0004 (0.001) | – 0.0002 (0.001) |
|     Travel | | – 0.001 (0.001) | – 0.001 (0.001) |
|     Other | | 0.001 (0.003) | 0.001 (0.003) |
| Constant | 0.004*** (0.001) | 0.003*** (0.001) | 0.002* (0.001) |
| Observations | 3,695 | 3,695 | 3,695 |
| R2 | 0.014 | 0.108 | 0.119 |
| Adjusted R2 | 0.011 | 0.104 | 0.112 |
| Residual Std. Error | 0.013 (df = 3681) | 0.013 (df = 3677) | 0.013 (df = 3664) |
| F Statistic | 4.024***(df = 3681) | 26.111*** (df = 3677) | 16.483*** (df = 3664) |
| *Controlled for: monetization; channel age* | | | * $p < .05$; ** $p < .01$; *** $p < .001$ |